\documentclass[preprint,aps]{revtex4}
\usepackage{graphics}

\usepackage{graphicx}
\usepackage{dcolumn}
\usepackage{bm}

\begin{document}


\title{Microwave screening by conduction currents in thin magnetic films: application in stripline broadband FMR}

\author{M.\ Kostylev}
\email{kostylev@cyllene.uwa.edu.au}
\author{K.~J.\ Kennewell}
\author{R. Magaraggia}
\author{R.~L.\ Stamps}
\affiliation{School of Physics, The University of Western Australia, 35 Stirling Highway, Crawley WA 6009, AUSTRALIA}
\author{M.\ Ali}
\author{D.\ Greig}
\author{B.~J.\ Hickey}
\affiliation{Department of Physics \& Astronomy, University of Leeds, Leeds, LS2 9JT United Kingdom}

\begin{abstract} 
Ferromagnetic resonance in conducting magnetic bilayers was studied using microstrip
transducers. It was found that excitation or suppression of standing spin waves could be achieved through enhanced inhomogeneity of eddy currents in the bilayer caused by finite thickness. This effect is observable in films with thicknesses below the magnetic skin depth and can be used to study standing spin wave modes in heterostructures.
\end{abstract}
\maketitle


Broadband inductive microwave spectrometers have become a common experimental tool with which to study dynamic properties of magnetic thin films and nano-structures \cite{Silva, Counil, Crew, self-organized}. Typically, resonant excitation and absorption is  accomplished using either a coplanar or microstrip microwave transmission line as the microwave transducer. The ferromagnetic sample is usually situated directly on top of the stripline. Absorption of microwave power by the sample loaded transmission line is conveniently measured using a network analyzer \cite{Counil} or through a combination of a  microwave generator and microwave passive components \cite{self-organized}. 

These techniques are useful for measuring accurately ferromagnetic resonance (FMR) and spin wave frequencies in the low GHz range in studies and characterizations of ferromagnetic structures.  As noted by several authors, careful consideration needs to be made of the particular experimental configuration when analyzing results \cite{Counil, Kim,  Schneider, Patton, Counil1, Schneider2}. In particular, Schneider et al. \cite{Schneider} mention that the measured response is dependent on the distance of the film from the coplanar surface due to stripline impedance. In another work they report the use of a floating ground plane above the film in order to enhance the response amplitude \cite{Schneider2}. Especially relevant for the present work, the broadband inductive technique can excite higher order spin wave resonances in addition to the fundamental resonance mode  \cite{Crew}. 

In the homogeneous microwave magnetic field of a microwave cavity, standing spin wave resonances (SSW) are observable provided there is a strong pinning of magnetization at the film surfaces (see Refs.\cite{Tannewald1, Tannewald2} and references therein). It is also known that effects of exchange interactions on ferromagnetic resonance in metals can be observed experimentally, provided the microwave field gradient is large (see discussion in Ref. \cite{Tannewald1}). Recently it was demonstrated theoretically \cite{kostylev} that conditions for a highly inhomogeneous microwave magnetic field are formed in the microstrip broadband FMR geometry due to microwave screening by eddy currents in conducting samples. As a result the microwave screening in conducting films thinner than the microwave magnetic skin depth \cite{Almeida} (hereby referred to as "sub skin depth films, SSDF") may strongly affect broadband FMR measurement results. In summary, manifestations of microwave screening for SSDF are as follows:

(i) response of multi-layered systems may strongly depend on layer ordering with respect to the microwave transducer location;    

(ii)  extremely large amplitudes of higher-order standing spin wave modes can be observed in some multi-layered systems;

and (iii) response of these systems can be strongly frequency dependent.

To some extent the driving of SSW discussed here is similar to efficient excitation of higher-order FMR modes by the microwave electric field \cite{Wolf}.  In his paper  Ref.\cite{Wolf} P.Wolf demonstrated excitation of SSW resonances in conducting films by a microwave electric field of potential nature. In Wolf's experiment the electric field applied in the film plane drives a microwave current in the sample. The highly inhomogeneous magnetic field of the current efficiently excites inhomogeneous resonances in the sample. In Ref.\cite{kostylev} it was noted that an external in-plane microwave magnetic field $\bf{h}_e$ applied to a film medium is necessarily accompanied by an in-plane curling electric field. This electric field should induce a microwave current in the sample whose Oersted field  $\bf{h}_{Oe}$ adds to the external microwave magnetic field. For thick samples one recovers the magnetic skin depth effect and the amplitude of the total field $\bf{h}_e+\bf{h}_{Oe}$ falls off exponentially with the distance from the sample surface facing the incident field flux. This thick-sample geometry has been discussed in the past \cite{Almeida, Ament}. Inside SSDF the total microwave magnetic field decays more strongly than exponential, such that for films with thicknesses larger than 30 nm  \cite{Permalloy} the field is negligible at the rear film surface. This results in a highly inhomogeneous and strongly asymmetric total microwave magnetic field inside the samples. 
Note that the inhomogeneity originates from extension of the classical skin depth effect to samples of finite thicknesses (see \cite{Bauer} and references therein).

Since  magnetization precession is driven by the total field, conditions are thereby formed for efficient excitation of non-uniform eigenmodes of precession. If eigenmodes of the system lack inversion symmetry the SSDF broadband FMR response will depend on layer ordering with respect to the direction of the incident microwave flux. This effect is useful for studying buried magnetic interfaces with FMR since observation of additional inhomogeneous resonances delivers unique information on interface properties. 

In the present work we illustrate the enhancement of SSW absorption with multilayer systems characterized by dynamic pinning of magnetization \cite{Wigen}. Here we used bi-layers comprised of a thin Cobalt layer and a much thicker Permalloy layer which are exchange coupled across the interface. This system's eigenmode structure is similar to calculated in Ref.\cite{kostylev}. The modes localized in the thin layer are shifted to very high frequencies and are not detected here. The observed dynamics are resonances localized in the thicker (Permalloy) layer (Fig. 1a-1b in \cite{kostylev}). 

Three series of samples were fabricated by magnetron sputtering at an argon working pressure of 2.5 mTorr.
The base pressure prior to the deposition was of the order of 10-8 Torr, and the films were deposited onto
silicon (100) substrates in an in-plane forming field of magnitude 200 Oe at ambient temperature.
Series one consisted of reference films of Ta(5 nm)/NiFe(30, 40, 60, 80 nm)/Ta(10 nm) where the Permollay
film thickness was varied as shown. The second ("Si/Py/Co")  and third ("Si/Co/Py")  series of samples were fabricated
as Ta(5 nm)/NiFe(30, 40, 60, 80nm)/ Co(10 nm)/Ta(10 nm)  and  Ta(5 nm)/Co(10 nm)/NiFe(30, 40, 60, 80nm)/Ta(10 nm) respectively, the difference between the two being the order of the two ferromagnetic layers. The total sample thickness was
chosen to be smaller in all samples in relation to the microwave magnetic skin depth in the frequency range
available with our microwave network analyzer (100 MHz-20 GHz) \cite{skin}.

In our experiments we use an Agilent N5230A PNA-L microwave network analyzer to apply a microwave signal to the samples and to measure magnetic absorption. As a measure of the absorption we use the microwave scattering parameter $S21$ \cite{Counil}. The driving current is applied to a section of microstrip line having a microstrip 1.5 mm in width. The sample sits on top of this microstrip transducer with the magnetic layers facing the microstrip. To avoid direct electric contact the sample surface is separated from the transducer by a teflon spacer 15 micron in thickness.  We keep the microwave frequency constant and sweep the static magnetic field $H$ applied in the film plane and along the transducer to produce resonance curves in the form of $S21(H)$ dependences. This is repeated for a number of frequencies. We also measure $S21$ for the transducer with no sample ($S21_0(H)$) to eliminate any field-dependent background signal from the results. The results presented below are $Re(S21(H)/S21_0(H))$. It is worth noting that the raw data $\mid S21(H) \mid $ show the same qualitative behavior, so artefacts arising from the mathematical processing of data are not significant.

Figure 1 contains representative experimental data, obtained for the microwave frequencies of 7.5 GHz and 18 GHz. 
Figures 1a and 1e contain the responses of the reference single-layer Py films 40 and 80nm thick. The position of the high-field intense peak is consistent with the fundamental mode. Since resonant fields for this mode practically coincide for these two Py films, there is no considerable surface anisotropy-related pinning of magnetization on the surfaces of Py. (The 30nm and 60nm thick reference films have the same resonant field also.) The smaller peak is consistent with the first (odd-symmetry) exchange mode for unpinned surface spins assuming values of the exchange constant of $A=6 \cdot 10^{-7}$ erg/cm, a gyromagnetic constant of 2.82 MHz/Oe, and the saturation magnetization for Permalloy $4 \pi M_s=$8320 G.

Data for bi-layers are shown in panels 1b to 1d and 1f to 1h. Each panel contains one response from a Si/Py/Co structure and one response from a Si/Co/Py structure with the same Py layer thickness. One sees that the response of Si/Co/Py structures is characterized by a single absorption peak located at a field which is a little smaller than that of the fundamental mode for the corresponding Py single-layer. We will call this single-peak response "Type A". The downshift decreases with increase in Py layer thickness which indicates dynamic magnetization pinning. The reversed ordering of layers (Si/Py/Co) shows quite a different response ("Type B"). Two peaks with comparable intensities are seen at 7.5 GHz (panels 1b-1d). The higher-field peak is located at the same field as the response of Type A, and thus it is consistent with the fundamental mode. The lower-field peak is at a field close the 1st SSW for the single-layer film and thus it is the first SSW mode for the structure. The thickness-dependent response is downshifted for this mode, with respect to the response for the single-layer films and is also a signature of dynamic pinning. The data contained in this figure are in full qualitative agreement with predictions in Ref.\cite{kostylev}. Features (i) to (iii) listed in the beginning of the present article are clearly seen. In particular, the experimental data for 18 GHz display a larger number of higher-order modes than for 7.5 GHz. Their intensities are extremely high, such that the first higher-order mode in panels 1f to 1h dominates in the resonant spectra.

In order to gain more evidence for the effect of screening by eddy currents additional experiments were performed. Following Appendix B, Case A in Ref. \cite{kostylev} the same effect can be seen for \textit{propagating} plane electromagnetic waves incident normally onto the film surface. We use a hollow metallic waveguide of rectangular cross-section to form conditions for the normal incidence. The films are in the plane of the waveguide's transverse cross-section. 
The measurements are made in reflection  \cite{half-cross-section}, so that the parameter S11 \cite{Counil} is obtained. Representative data are contained in Fig. 2. 
We find the same tendency as in the microstrip experiments: Responses of Type A are obtained when the Co-layer of any structure faces the incident flux and responses of Type B are obtained when the Py layer faces the incident flux. Furthermore, when the samples are placed with substrates facing the incident flux, the samples which previously showed Type A responses show responses of type B, and vice versa \cite{footnote substrate}. 

In conclusion, we find strong variation in stripline response depending on the ordering of layers of bi-layer structures with respect to a stripline transducer.  All experimental data are consistent with predicted manifestations of the effect of screening by eddy currents in films with thicknesses below the microwave magnetic skin depth. This effect may be useful for studying buried magnetic interfaces.

We acknowledge support from the Australian Research Council and the University of Western Australia.

\newpage
Figure captions:

Fig. 1. Microwave stripline broadband FMR absorption. Microstrip transducer is 1.5 mm in width. Left-hand panels (a-d): driving frequency is 7.5 GHz. Right-hand panels (e-h): 18 GHz.

(a) and (e): Single-layer Permalloy films (Si/Ta/Py/Ta); solid line: 80 nm thick; dashed line: 40 nm thick;

(b-d) and (f-h) are bi-layer responses. Red solid lines are for Si/Ta/Py/Co/Ta and dashed blue lines are for Si/Ta/Co/Py/Ta structures. In all figures Cobalt layer is 10nm thick. 

(b) and (f): Permalloy thickness is 40 nm;
(c) and (g): 60 nm;
(d) and (h): 80 nm.

Fig. 2. Hollow waveguide data. Frequency is 9.5 GHz. (a): response of Si/Ta/Py[80nm]/Co[10 nm]/Ta structure. (b): response of Si/Ta/Co[10 nm]/Py[80 nm]/Ta structure. Red solid line: Film facing the incident flux. Blue dashed line: substrate facing the incident flux. All measurements were done in reflection (S11 parameter was measured).

\begin{figure}
	\centering
		\includegraphics
{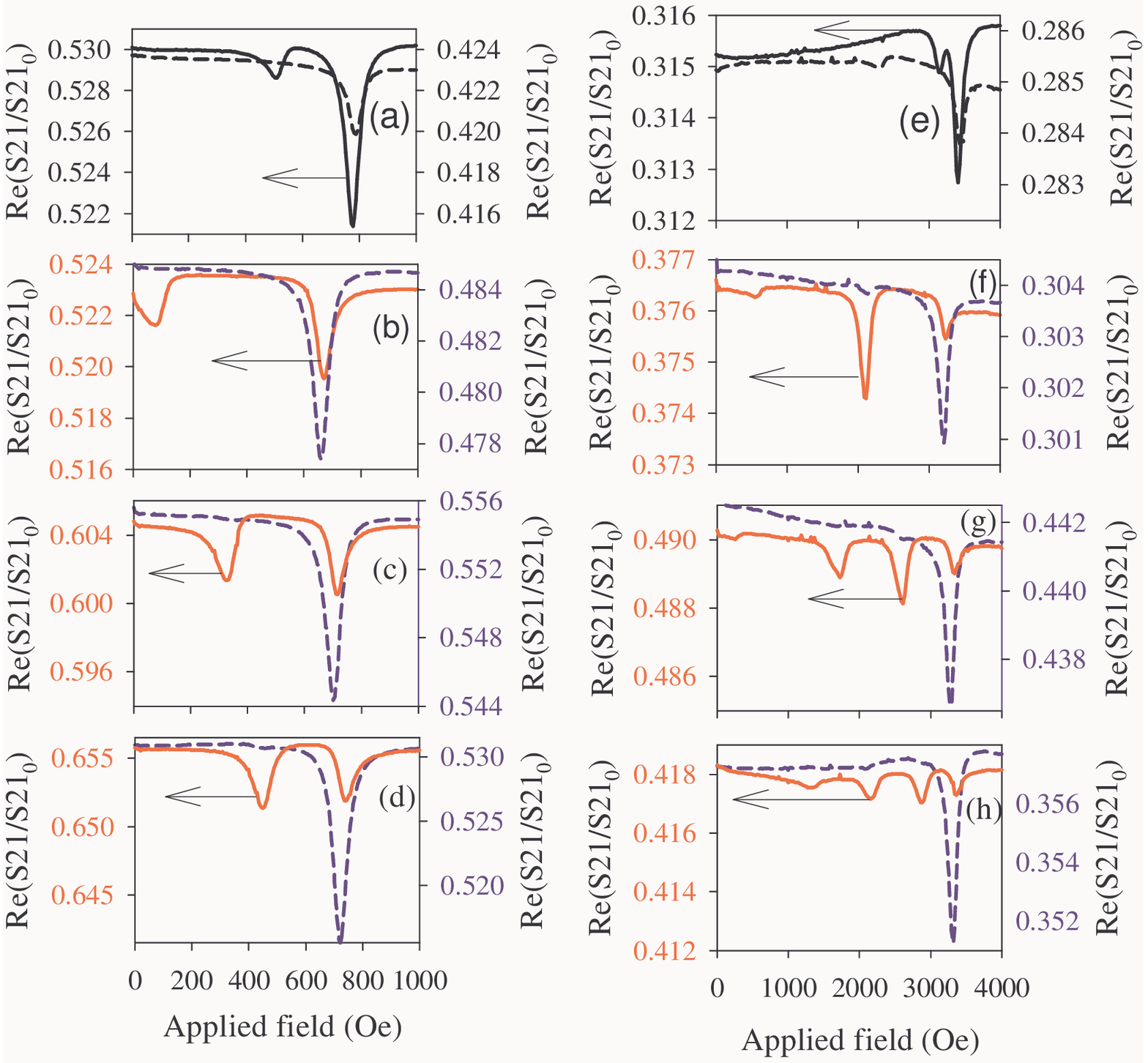}
\end{figure}

\newpage

\begin{figure}
	\centering
		\includegraphics{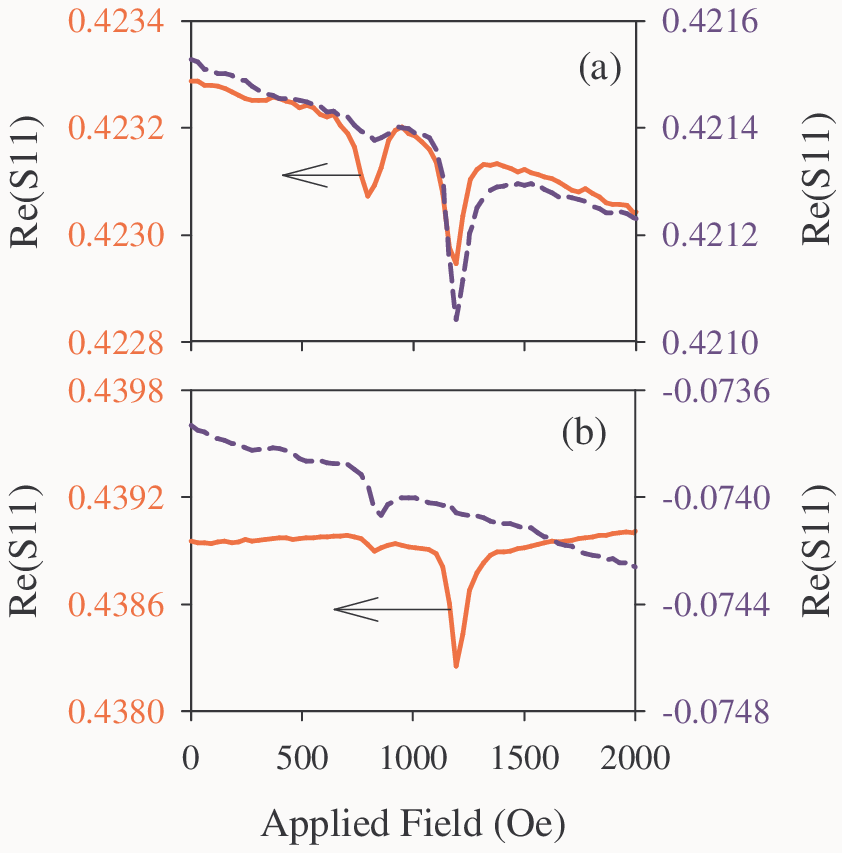}
\end{figure}




\begin{thebibliography}{sotief}
\bibitem{half-cross-section} The samples fill about one half of the the waveguide cross-section. For this reason one has to expect a microwave field incident at the real sample boundary due to wave diffraction through the waveguide cross-section not filled with the sample. The presence of the diffracted field is confirmed by our network analyzer measurements. When a sample is inserted the waveguide transmission characteristic $S21(f)$ becomes non-monotonous with frequency $f$ because of partial standing wave resonances formed in the waveguide. Depending on frequency  $S21$ varies from -6db to -12dB. Our field resolved measurements are carried out in a local minimum of $S21(f)$ in order to minimize incidence of the diffracted microwave field at the rear sample surface.
\bibitem{footnote substrate} Swapping from Type A to Type B responses and vice versa has been also observed when putting samples on the microstrip with substrates facing the transducer.
\end{thebibliography}
\end{document}